\documentstyle[preprint,aps,eqsecnum,epsfig]{revtex}

\renewcommand{\baselinestretch}{1.13}

\newcommand{\lbar}{\overline}
\newcommand{\spa}{\vspace{.25cm}}
\newcommand{\beq}{\begin{equation}}
\newcommand{\eeq}{\end{equation}}
\newcommand{\beqs}{\begin{eqnarray}}
\newcommand{\eeqs}{\end{eqnarray}}

\newcommand{\eV}{\mbox{\,{\rm eV}}}

\newcommand{\bM}{\mbox{\boldmath $M$}}
\newcommand{\bV}{\mbox{\boldmath $V$}}

\newcommand{\phii}{i}

\newcommand{\SUSYwoR}{SUSY $\not\!\!R_p$}

\def\lsim{\ \rlap{\raise 3pt \hbox{$<$}}{\lower 3pt \hbox{$\sim$}}\ }
\def\gsim{\ \rlap{\raise 3pt \hbox{$>$}}{\lower 3pt \hbox{$\sim$}}\ }

\def\vev#1{\langle #1 \rangle}
\def\half{{\textstyle{1 \over 2}}}
\def\eighth{{\textstyle{1 \over 8}}}

\def\npb#1{Nucl.\ Phys.\ {\bf B#1}}
\def\plb#1{Phys.\ Lett.\ {\bf B#1}}
\def\prd#1{Phys.\ Rev.\ {\bf D#1}}
\def\prl#1{Phys.\ Rev.\ Lett. {\bf#1}}

\def\zpc#1{Z.~Phys.\ {\bf C#1}}
\def\epjc#1{Eur.~Phys.~J.\ {\bf C#1}}

\def\mpla#1{Mod. Phys. Lett. A {\bf #1}}
\def\JHEP#1{JHEP {\bf #1}}

\begin{document}

{\tighten %
\preprint{\vbox{\hbox{WIS-4/00/Apr-DPP}
                \hbox{hep-ph/0004048}
                }}

\title{~ \\ Lepton number violation interactions \\
            and their effects on neutrino oscillation
            experiments}
\author{Sven Bergmann\,$^a$,
H.V. Klapdor--Kleingrothaus\,$^b$ and Heinrich P\"as\,$^b$}
\address{ \vbox{\vskip 0.truecm}
  $^a$Department of Particle Physics \\
  Weizmann Institute of Science, Rehovot 76100, Israel \\
\vbox{\vskip 0.truecm}
  $^b$Max--Planck--Institut f\"ur Kernphysik\\
P.O. Box 103980, D-69029 Heidelberg, Germany}

\maketitle

\begin{abstract}
  Mixing between bosons that transform differently under the standard
  model gauge group, but identically under its unbroken subgroup, can
  induce interactions that violate the total lepton number.  We
  discuss four-fermion operators that mediate lepton number violating
  neutrino interactions both in a model-independent framework and
  within supersymmetry (SUSY) without $R$-parity.  The effective
  couplings of such operators are constrained by: i)~the upper bounds
  on the relevant elementary couplings between the bosons and the
  fermions, ii)~by the limit on universality violation in pion decays,
  iii)~by the data on neutrinoless double beta decay and, iv)~by
  loop-induced neutrino masses. We find that the present bounds imply
  that lepton number violating neutrino interactions are not relevant
  for the solar and atmospheric neutrino problems. Within SUSY without
  $R$-parity also the LSND anomaly cannot be explained by such
  interactions, but one cannot rule out an effect model-independently.
  Possible consequences for future terrestrial neutrino oscillation
  experiments and for neutrinos from a supernova are discussed.
\end{abstract} 
} 

\newpage

\section{Introduction}

Experimentalists have reported three different kinds of ``neutrino
anomalies'', which seem to indicate that the standard model (SM)
description of the neutrino is incorrect.  Today, many physicists
consider the recent SuperKamiokande high statistics
result~\cite{SKAN}, which confirmed the long-standing atmospheric
neutrino (AN) problem~\cite{ANP}, as the strongest experimental
evidence for New Physics (NP) beyond the SM. However, also
increasingly convincing arguments, that the solar neutrino (SN) data
can only be explained by extending the SM neutrino picture, have been
established in recent years~\cite{SNP}. Finally the LSND collaboration
has found unexpected signals for neutrino flavor conversion in two
appearance experiments~\cite{LSND-DAR,LSND-DIF}. So far none of the
other short baseline experiments~\cite{KARMEN,Bugey,CHOOZ}, has been
able to confirm these results, but major experimental efforts are
underway to search for neutrino oscillations both at
short~\cite{MiniBooNE} and long baseline~\cite{Zuber,K2K,MINOS,ICANOE}
facilities.

The favorite explanation for the existing neutrino anomalies is to
allow for massive neutrinos that mix and therefore undergo flavor
oscillations while propagating. Neutrino oscillations provide
convincing solutions to each of the above mentioned neutrino problems.
However, the SN, the AN and the LSND observations imply three
separated scales for the mass-squared
differences~$\Delta_{ij}=m^2_i-m^2_j$
\beqs
\Delta_{SN} &\lsim& 10^{-5} \eV^2 \,,   \label{DeltaSN} \\
\Delta_{AN} &\sim& 10^{-3} \eV^2 \,,    \label{DeltaAN} \\
\Delta_{LSND} &\gsim& 10^{-1} \eV^2 \,, \label{DeltaLSND}
\eeqs
which cannot be accommodated simultaneously in a three neutrino
framework~\cite{Giunti}.  Consequently, unless one ignores one of the
three anomalies or allows for a forth non-sequential light
neutrino~\cite{sterile}, already the present neutrino data indicate
that neutrino masses and mixing alone might not be the complete
picture of the New Physics in the neutrino sector. It is important to
note that many extensions of the SM that could provide massive
neutrinos also predict non-standard neutrino interactions. In fact, in
some cases new interactions induce neutrino masses in loop
processes~\cite{mass-ind} and one can relate the two aspects of New
Physics quantitatively.

Solutions of the various neutrino anomalies in terms of new
interactions with and without neutrino masses and mixing have been
studied in Ref.~\cite{SN-NP,AN-NP,Grossman,Herczeg,Johnson,Bergmann,Bea}.
While for the SN problem this is indeed a viable
possibility~\cite{Bergmann,Bea} it has been shown that new flavor
changing neutrino interactions that conserve total lepton number are
constrained by the high precision data that confirm the SM predictions
to be too small to affect the atmospheric~\cite{BGP} and the
LSND~\cite{BG} anomalies.

In this work we study another class of NP interactions, which so far
has only received little attention~\cite{Herczeg}, namely new neutrino
interactions that violate the {total lepton number}~$L$.  Such
interactions can arise naturally in models where there is mixing
between bosons that transform differently under the SM gauge group,
but identically under its unbroken subgroup.  As an example consider
the anomalous muon decays that produce two antineutrinos
\beq \label{amdlv}
\mu^+ \to e^+ \, \bar\nu_e \, \bar\nu_\ell \qquad (\ell=e, \mu, \tau).
\eeq
Such decays violate $L$ by two units.  In principle the reaction
in~(\ref{amdlv}) could produce the $\bar\nu_e$'s that are observed at
LSND in the decay at rest (DAR) channel and which are usually accounted for by
$\bar\nu_\mu \to \bar\nu_e$ flavor oscillations.

Unlike for the $L$-conserving interactions, replacing the
antineutrinos by their (positively) charged $SU(2)_L$ partners gives
rise to interactions that violate $U(1)_{EM}$.  Thus, it follows that
the effective couplings of the above decays (\ref{amdlv}) must vanish
in the $SU(2)_L$ symmetric limit and be proportional to $SU(2)_L$
breaking effects. This breaking is not proportional to a mass
splitting within a given multiplet as for the $L$-conserving
interactions~\cite{BG}, but it shows up as {\it mixing} between
(heavy) bosons of different $SU(2)_L$ representations.

In Section~\ref{formalism} we present the general framework that
expresses the effective strength of the lepton number violating
interactions (as well as those that conserve total lepton number) in
terms of the boson masses, their mixing angle and the relevant
trilinear couplings.  In Section~\ref{models} we discuss supersymmetry
without $R$-parity (\SUSYwoR) as a prominent example for such a
scenario, where the mixing between left-handed and right-handed
sfermions that couple to the SM fermions via $R_p$ violating
interactions, induces lepton number violating interactions.  In
Section~\ref{constraints} we establish relations between lepton number
violating interactions and those that conserve total lepton number.
We use these relations to derive constraints on the new interactions.
Additional bounds on these interactions arise from the limit on
universality violation in pion decays, the data on neutrinoless double
beta decay and from loop-induced neutrino masses.  In
Section~\ref{exps} we investigate whether the lepton number violating
interactions could be relevant for any of the three anomalies as well
as for the up-coming terrestrial neutrino oscillation experiments.
Also implications for neutrinos from a supernova are discussed.  We
conclude in Section~\ref{conclusions}.

\section{Formalism}
\label{formalism}

Consider a generic extension of the standard model with two bosonic
fields $\phi$ and $\chi$ that transform differently under $SU(2)_L$.
In general, after $SU(2)_L$ breaking, the $SU(2)_L$ components of
these fields $\phi_q$, $\chi_q$ which transform identically under the
unbroken SM gauge group $SU(3)_C \times U(1)_{EM}$ can mix with each
other giving rise to a hermitian mass-matrix
\beq \label{massmatrix}
\bM^2 = \pmatrix{ M_{11}^2 & M_{12}^2 \cr
                  M_{21}^2 & M_{22}^2 \cr} \,.
\eeq
Diagonalizing $\bM^2$ yields the eigenvalues
\beq \label{eigenvalues}
M_{1,2}^2 = 
{1 \over 2} \left(\Sigma \mp \sqrt{\delta^2 + 4|M_{12}^2|^2} \right) \,,
\eeq
with $\Sigma=M_{11}^2 + M_{22}^2$ and $\delta^2=(M_{11}^2 -
M_{22}^2)^2$.  The mass-eigenstates are linear combinations of $\phi_q$
and $\chi_q$, i.e.
\beq
|\phii\bigr> = V_{i1} |\phi_q\bigr> + V_{i2} |\chi_q \bigr> \,,
\qquad (i=1, 2) \,.
\eeq
Assuming that $M_{12} = M_{21}$ is real, the mixing matrix can be
parameterized as
\beq \label{mix}
\bV = \pmatrix{ \cos\theta &  \sin\theta \cr
               -\sin\theta &  \cos\theta \cr} \,,
\eeq
with
\beq \label{mixing}
\sin2\theta = {2 M_{12}^2 \over \sqrt {\delta^2 + 4 M_{12}^4}}.
\eeq
Let us add now renormalizable interactions that couple the bosonic
fields $\phi$ and $\chi$ to bilinears $A$, $B$ which are built out of
two SM fermions
\beq \label{trilinear} 
-{\cal L}_{A,B} =
\lambda_A \, (\phi \, A) + \lambda_B \, (\chi \, B) + \rm{h.c.} \,, 
\eeq
where $\lambda_A$ and $\lambda_B$ denote the elementary trilinear
couplings. These couplings are induced by New Physics that may be
present at or above the weak scale. Any such theory will include the
SM gauge symmetry, implying that ${\cal L}_{A,B}$ is invariant under
${\cal G}_{\rm SM}=SU(3)_C \times SU(2)_L \times U(1)_Y$. If the
bosons are vector fields the couplings in~(\ref{trilinear}) may be
gauge interactions. We will focus on scalar bosons that couple to
fermions via {\it a priori} arbitrary Yukawa couplings
$\lambda_{A,B}$.  The fermion bilinears may be composed of quarks,
leptons or both, as well as the respective antiparticles, which all
belong to either $SU(2)_L$ singlets or doublets.  Then gauge
invariance implies that the bosonic fields may be singlets (s),
doublets (d) or triplets (t) of $SU(2)_L$.  Since we require $\phi$
and $\chi$ to have different transformation properties under $SU(2)_L$
also $A$ and $B$ will transform differently. On the other hand, since
$\phi$ and $\chi$ transform identically under $SU(3)_C$ this also
applies to $A$ and $B$.

Given the elementary couplings in~(\ref{trilinear}) one can construct
four-fermion interactions that are mediated by the bosonic fields.
Each bilinear $A$ and $B$ can be either coupled to itself or there can
be a coupling {\it between} $A$ and $B$. Since $\phi_q$ and $\chi_q$
are required to have the same electric charge $q$ there is only a
coupling between the $SU(2)_L$ components $A_q$ and $B_q$ that have
the same charge such that the resulting four-fermion operator
$A_q^\dagger B_q$ conserves $U(1)_{EM}$.  Similarly, since $A$ and $B$
transform identically under $SU(3)_C$ it follows that $A_q^\dagger
B_q$ is a color singlet.

We stress that the coupling between fermion bilinears that have
different $SU(2)_L$ transformations requires the mixing between
$\phi_q$ and $\chi_q$. As a unique consequence the coupling between a
bilinear that transforms as an $SU(2)_L$ doublet to a bilinear that is
a singlet or a triplet of $SU(2)_L$ can induce effective four-fermion
operators that do not conserve the total lepton number $L$.  In
addition such operators may also violate the individual lepton number
$L_\ell$.  For example, if the $SU(2)_L$ doublet $A=\bar L_e E_\mu$
($E_\ell$ and $L_\ell$ denote a lepton singlet and doublet field,
respectively, of flavor $\ell$) couples to an $SU(2)_L$ doublet $\phi$
and if the $SU(2)_L$ singlet $B_{-1}=(L_\mu L_e)_s$ couples to an
$SU(2)_L$ singlet $\chi^+$, then the mixing between the $q=1$
doublet-component $\phi^+$ and $\chi^+$ gives rise to the operator
$A_{-1}^\dagger B_{-1} = (\lbar{\mu_R} \nu_e) \, (\nu_\mu e_L - \mu_L
\nu_e)$, which induces $\mu^+_L \to e^+_R \, \bar\nu_\mu \, \bar\nu_e$
and $\mu^+_L \, \nu_e \to \mu^+_R \bar\nu_e$. Both processes violate
$L_e, L_\mu$ and $L$ by two units.

Note that if one scalar field couples to two different bilinears
(which, consequently must have the same $SU(2)_L$ transformation) then
these two bilinears can be coupled to each other.  However, the four
fermion operators that arise from this mechanism (see e.g.~\cite{BG})
may only violate $L_\ell$, but not $L$.

The effective four-fermion operators $A_q^\dagger A_q$, $B_q^\dagger
B_q$ and $A_q^\dagger B_q$ at energies well below the masses of the
scalar fields [i.e. the eigenvalues of $\bM^2$ given
in~(\ref{eigenvalues})] are obtained by integrating out the bosonic
degrees of freedom. Assuming weak trilinear couplings, $\lambda_{A,B}
\lsim 1$, the tree-level diagrams result into the effective couplings
\beq
G_N^{A^\dagger A} = {|\lambda_A|^2 \over 4 \sqrt2 M_A^2}, \qquad
G_N^{B^\dagger B} = {|\lambda_B|^2 \over 4 \sqrt2 M_B^2}, \qquad
G_N^{A^\dagger B} = {\lambda_A^* \lambda_B \over 4 \sqrt2 M_{AB}^2}, 
\eeq
where the respective low-energy propagators are given by
\beqs \label{propagators}
M_A^{-2} &\equiv& 
\sum_i {\bigl<\phi|\phii\bigr> \bigl<\phii|\phi\bigr> \over M_i^2} =  
{\cos^2\theta \over M_1^2} + {\sin^2\theta \over M_2^2} \,,  \\
M_B^{-2} &\equiv& 
\sum_i {\bigl<\chi|\phii\bigr> \bigl<\phii|\chi\bigr> \over M_i^2} =  
{\cos^2\theta \over M_2^2} + {\sin^2\theta \over M_1^2} \,, \\
\label{MAB}
M_{AB}^{-2} &\equiv& \label{propagatorAB}
\sum_i {\bigl<\phi|\phii\bigr> \bigl<\phii|\chi\bigr> \over M_i^2} = 
{\sin2\theta \over 2} \left({1 \over M_1^2} - {1 \over M_2^2} \right). 
\eeqs
From~(\ref{MAB}) it is obvious that $A_q^\dagger$ and $B_q$ can only
couple to each other when there is a non-vanishing mixing
($\sin2\theta \ne 0$) and when the physical masses are not degenerate
($M_1 \ne M_2$).  We note that for maximal mixing ($\sin2\theta = 1$)
the propagator $M_A^{-2}$ equals to $M_B^{-2}$, and $M_{AB}^{-2}$ is
maximal.  Moreover, we remark that using~(\ref{mixing}) it follows
that the propagator in~(\ref{MAB}) is simply
\beq \label{MAB2}
M_{AB}^{-2} = {M_{12}^2 \over M_1^2 \, M_2^2} \,.
\eeq

\section{Lepton number violation in SUSY without $R$-parity}
\label{models}

In this section we present an explicit example for the general
mechanism developed in Section~\ref{formalism} by discussing scalar
mixing in supersymmetric extensions of the standard model without
$R$-parity~\cite{SUSY}.  In particular, we show how the model-specific
parameters of this theory translate into those we introduced in
Section~\ref{formalism}.

The $R$-parity violating couplings 
$\lambda_{\imath \jmath \kappa} L_\imath L_\jmath E_\kappa^c$ and
$\lambda'_{\imath \jmath \kappa} L_\imath Q_\jmath D_\kappa^c$,
where $L_\kappa, Q_\kappa, E_\kappa$ and $D_\kappa$ denote the chiral
superfields containing, respectively, the left-handed lepton and quark
doublets and the right-handed charged-lepton and $d$-quark singlets of
generation $\kappa=1, 2, 3$, introduce a variety of couplings between
fermion bilinears and sfermions:
\beqs
-{\cal L}_{\lambda} &=& {\lambda_{\imath \jmath \kappa} }
\left[\tilde\nu_L^\imath \lbar{\ell_R^\kappa} \ell_L^\jmath 
+\tilde\ell_L^\jmath \lbar{\ell_R^\kappa} \nu_L^\imath
+\tilde\ell_R^{\kappa *} \lbar{\nu_L^\imath}^c \ell_L^\jmath 
-(\imath \to \jmath )\right]+{\rm h.c.} \, ,
\label{lepSS}
\\
-{\cal L}_{\lambda'} &=& {\lambda'_{\imath \jmath \kappa} }
\left[\tilde\nu_L^\imath \lbar{d_R^\kappa} d_L^\jmath 
+\tilde d_L^\jmath \lbar{d_R^\kappa} \nu_L^\imath
+\tilde d_R^{\kappa *} \lbar{\nu_L^\imath}^c d_L^\jmath 
-\tilde e_L^\imath \lbar{d_R^\kappa} u_L^\jmath 
-\tilde u_L^\jmath \lbar{d_R^\kappa} e_L^\imath
-\tilde d_R^{\kappa *} \lbar{e_L^\imath}^c u_L^\jmath 
\right]+{\rm h.c.} \, .
\label{shSS}
\eeqs
Due to charge conservation only sfermions of the same type can mix.
In principle mixing is allowed between the left-handed and the
right-handed components of the (charged) sfermions, as well as between
sfermions of different generation, but for simplicity we will assume
that the latter is negligible.

The leptonic couplings in~(\ref{lepSS}) can induce $L$-violating
interactions like in~(\ref{amdlv}) \cite{Herczeg}. 
For example, identifying the
scalar fields $\phi^+=\tilde\tau_R^+$ and $\chi^+=\tilde\tau_L^+$ and
the couplings $\lambda_A=\lambda_{132}^*$ and
$\lambda_B=\lambda_{123}$ reproduces exactly the example in
Section~\ref{formalism} that gave rise to $\mu^+_L \to e^+_R \,
\bar\nu_\mu \, \bar\nu_e$. Note that SUSY without $R_p$ not only
provides the required scalar fields and their couplings, but also an
explicit expression for the mass-matrix in (\ref{massmatrix}),
i.e.~\cite{SUSYrev}
\beq \label{massmatrixSS}
{\bM_{\!\!\tilde f}}^{\!2} =
\pmatrix{ M_{\tilde L}^2 + m_f^2 + M_Z^2 (T_3^f/2 - q_f \sin^2\theta_W) 
        & m_f (A_f - \mu \cot^{T_3^f}\beta)  \cr
          m_f (A_f - \mu \cot^{T_3^f}\beta) 
        & M_{\tilde R}^2 + m_f^2 + M_Z^2 \, q_f \sin^2\theta_W},
\eeq
where $m_f$ and $q_f$ denote the mass and the charge of the fermion
$f$, $T_3^f=1~(-1)$ for $f=u_\kappa~(d_\kappa, e_\kappa)$, $M_{\tilde
L}^2$ $(M_{\tilde R}^2)$ is the soft supersymmetric breaking
mass-squared term for the left- (right-)handed sfermion, and $A_\ell$,
$\mu$ and $\tan\beta$ are the familiar SUSY parameters~\cite{SUSYrev}.

We note that in the absence of right-handed (s)neutrinos $\tilde
f_L-\tilde f_R$ mixing can occur only for charged sfermions. This
implies that the mass-matrix~(\ref{massmatrixSS}) has to be positive
definite in order to avoid the spontaneous breaking of $U(1)_{EM}$.

\section{Experimental Constraints}
\label{constraints}

In this section we discuss constraints on the effective couplings for
lepton number violating neutrino interactions.  As we mentioned
already, the corresponding four-fermion operators cannot be related to
the ones where the neutrinos are rotated into their charged lepton
partners, since such an $SU(2)_L$ rotation violates $U(1)_{EM}$.
Hence, while in many cases such a rotation can provide stringent
bounds on the product of trilinear couplings for interactions that
only violate $L_\ell$ (see~\cite{BGP,BG,Herczeg}), it does not help for
$L$-violating neutrino interactions. Instead one can use the
constraints on each of the trilinear couplings which arise from the
interactions induced by the self-couplings of a specific fermion
bilinear relevant for the lepton number violating neutrino
interaction.  Alternatively, in some cases, there are direct
constraints on the $L$-violating interactions.  Operators that induce
lepton number violating pion decays can be constrained using the limit
on universality violation.  Upper bounds on certain operators
containing the electron neutrino follow from the data on neutrinoless
double beta decay. Finally, in case the $L$-violating operator
involves two neutrinos, one can connect the two external charged
fermions in order to generate neutrino masses and use their upper
bounds.

\subsection{Constraints from the trilinear couplings}

Any non-vanishing trilinear coupling $\lambda_A$ between a fermion
bilinear $A$ and a boson $\phi$ can be used to create the effective
interaction
\beq
{|\lambda_A|^2 A^\dagger A \over 4 \sqrt2 M_A^2} \,.
\eeq
If the intermediate boson does not mix, the low-energy propagator is
simply $M_A^{-2}=M_\phi^{-2}$, but if there is mixing the correct
expression is the one in~(\ref{propagators}).

This is important, since the constraints on various trilinear
couplings in the literature are derived assuming that the respective
intermediate boson is a mass eigenstates which has a definite mass
$M$.  Therefore, if we denote the upper bound on any trilinear
coupling derived under such an assumption by $\hat\lambda_A$, then
this implies for the parameter $\lambda_A$, which describes the
coupling between $A$ and a boson $\phi$ that is not a mass eigenstate
(but which mixes with a different boson $\chi$ as discussed in
Section~\ref{formalism}), that
\beq
|\lambda_A| < \hat\lambda_A \times {M_A \over M}.
\eeq
This rescaling corrects for the fact that if the effective propagator
$M_A^{-2}$ is smaller (larger) than $M^{-2}$, then the constraint on
$\lambda_A$ will be weaker (stronger).

\begin{figure}[htb]
\LARGE
$\varrho$
\begin{center}
\mbox{\epsfig{figure=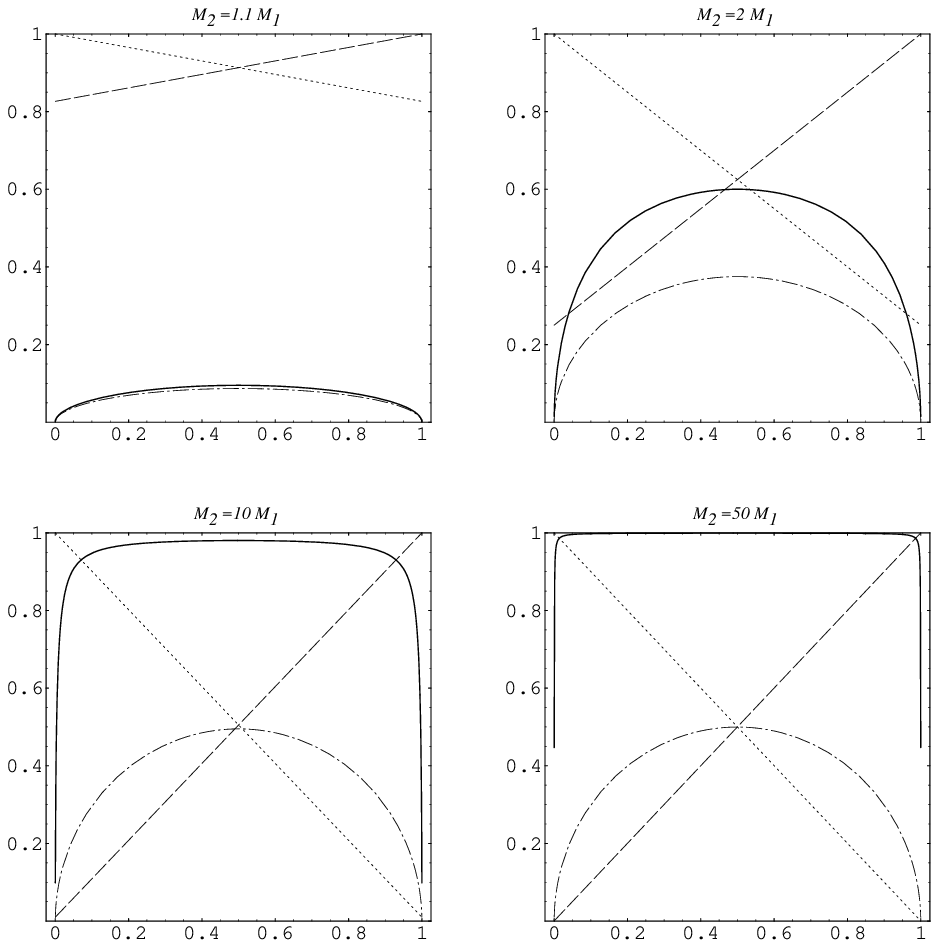,angle=0,width=16.6cm,height=16.2cm}}
\newline
\mbox{~~~~}$\sin^2\theta$
\spa
\end{center}

\caption{The solid curve shows the suppression factor $\varrho = {M_A
\, M_B \over M_{AB}^2}$ (horizontal axes) as a function of the
$\phi-\chi$ mixing $\sin^2\theta$ (vertical axes) for $M_2/M_1=1.1, 2,
10, 50$. Also the dependence on $\sin^2\theta$ for the propagators
$M_A^{-2}, M_B^{-2}$ and $M_{AB}^{-2}$ (in units of $M_1^-2$) is
indicated by the dotted, dashed and dashed-dotted curves,
respectively.}

\label{rho}
\end{figure}

Consequently the upper bound on any $L$-violating operator
$(G_N^{A^\dagger B}/\sqrt{2}) \, A^\dagger B$ which is induced by
$\phi-\chi$ mixing is constrained by
\beq \label{GABc}
G_N^{A^\dagger B} = {\lambda_A^* \lambda_B \over 4 \sqrt2 M_{AB}^2}
< {\hat\lambda_A^* \hat\lambda_B \over 4 \sqrt2 M^2} \times 
{M_A \, M_B \over M_{AB}^2}.
\eeq
The upper bound on the right-hand side of (\ref{GABc}) factorizes into
\beq
\hat G_N^{A^\dagger B} \equiv  
{\hat\lambda_A^* \hat\lambda_B \over 4 \sqrt2 M^2},
\eeq
which only depends on the upper bounds $\hat\lambda_A$ and
$\hat\lambda_B$ derived from experimental observations (under the
assumption that the intermediate particle has mass $M$) and the ratio
\beq
\varrho \equiv {M_A \, M_B \over M_{AB}^2}
\eeq
which is a function of the mixing angle $\theta$ and the mass
eigenvalues $M_1$ and $M_2$ only. Note that $\varrho(\sin \theta, M_1,
M_2) \le 1$ and that $\varrho$ is maximal at $\sin\theta =
\cos\theta=1/\sqrt2$, where it takes the value $\hat\varrho =
(M_2^2-M_1^2)/(M_2^2+M_1^2)$, which is small when the masses are
almost degenerate, but it quickly approaches unity when the degeneracy
is lifted.  We show $\varrho(\sin^2\theta)$ for various values of
$M_2/M_1$ in Fig.~\ref{rho}.

Since the interactions induced by the self-couplings of any fermion
bilinear $A$ do not violate $L_\ell$ and $L$, the corresponding NP 
operator only induces additional contributions to reactions that are
already present in the standard model. Therefore any non-zero NP 
effective coupling $G_N^{A^\dagger A}$ modifies the SM predictions for
the relevant processes and precision measurements can be used to put
upper bounds on $G_N^{A^\dagger A}$.

It is conventional to assume that only one trilinear coupling
$\lambda_A$ is non-zero for each bound and that the intermediate boson
has a mass of $M=100$~GeV. Then the constraint is expressed in terms
the dimensionless real number $\hat \lambda_A$ as
\beq
|\lambda_A| < \hat \lambda_A \times \left(M \over 100~\rm{GeV} \right),
\eeq
which translates into
\beq
G_N^{A^\dagger A} < \hat G_N^{A^\dagger A} \equiv 
{\hat \lambda_A^2 \over 4\sqrt{2}(100~\rm{GeV})^2} = 
1.52 \, \hat \lambda_A^2~G_F.
\eeq
for the effective coupling. If the process that was used to derive the
bound and the one which one wants to constrain are mediated by bosons
which are different members of the same $SU(2)_L$ multiplet, then one
has to correct for differences in the propagators
\beq
G_N^{q'} = {M_q^2 \over M_{q'}^2} \, G_N^{q}
< 1.52 \, \hat\lambda^2~{M_q^2 \over M_{q'}^2}~G_F \,,
\eeq
where $q, q'$ refer to the charge of the intermediate boson.  In
Ref.~\cite{BGP} it has been shown that electroweak precision data
imply that $M_q / M_{q'}$ is of order unity. Even for masses close to
the weak scale this ratio is at most $2.6$ unless one allows for some
fine-tuned cancellations.

In Tab.~1 we list all bilinears that couple to scalar weak singlets or
doublets that appear in SUSY without $R$-parity.  We also show in
Tab.~1 the upper bounds (at $2\sigma$) for both the trilinear
couplings ($\hat \lambda$) and the effective couplings ($\hat
G_N^{A^\dagger A}$). We assume here that $M_q / M_{q'} = 1$, bearing
in mind that for scalar doublets the maximal correction from $SU(2)_L$
breaking effects could be a factor a few.  For the bounds we use the
results~\cite{RpBounds} obtained within the framework of \SUSYwoR. All
limits are at $2\sigma$, except for $\lambda_{1\kappa1}$ which is at
$3\sigma$. The most stringent constraints relevant to our discussion
arise from charged current universality ($V_{ud}$), lepton
universality [$R_\tau=\Gamma(\tau \to e \nu \bar\nu) / \Gamma(\tau \to
\mu \nu \bar\nu)$, $R_\pi=\Gamma(\pi \to e \nu) / \Gamma(\pi \to \mu
\nu)$ and $R_{\tau \pi}=\Gamma(\tau \to \pi \nu_\tau) / \Gamma(\pi \to
\mu \nu_\mu)$], forward-backward asymmetries in $e^+ e^-$ collisions
at the $Z$ peak ($A_{FB}$), atomic parity violation (APV), $\nu_\mu$
deep inelastic scattering ($\nu_\mu$ DIS) and constraints on the
compositeness scale [$\Lambda(qqqq)$].  We note that the listed bounds
apply to any theory that contains the respective trilinear coupling,
since we consider only the constraints that are derived directly from
a specific coupling, that is we allow only one term in the
$R_p$-violating couplings~(\ref{lepSS}) and~(\ref{shSS}), to be
non-zero at a time. If one relaxes this assumption and takes all the
$R_p$-violating couplings together as they appear in~(\ref{lepSS})
and~(\ref{shSS}), i.e. one evokes supersymmetry, then in some cases
additional constraints (given is square brackets) can be found from
processes which have different intermediate scalars, but rely on the
same trilinear coupling.  We note that for the coupling $\lambda_{31}$
of $L_3 \lbar D_1$ to a scalar doublet to the best of our knowledge
there is no model-independent bound. Demanding that the theory remains
perturbative at large energies implies that $\lambda_{31} \lsim 1$. In
\SUSYwoR\ $\lambda_{31}=\lambda'_{3\kappa1} < 0.52$, which is due to
the upper bound on $\lambda'_{321}$ from $D_s$ decays.

\begin{center} 
Tab.~1: Experimental constraints on fermion bilinear self-couplings \\
\spa
\spa
\begin{tabular}{|c|c|c|c|c|c|}
\hline  
$\lambda_A$ & $A$ & $\phi$ & ~$\hat \lambda_A$ [\SUSYwoR]~ & 
~$\hat G_N^{A^\dagger A}/G_F$ [\SUSYwoR]~ & from \\
\hline \hline  
$\lambda_{12\kappa}$   & $L_1 L_2$       & ~$\tilde e_R^\kappa$~ & 
0.05 & 0.0038 & $V_{ud}$ \\ 
\hline 
$\lambda_{13\kappa}$   & $L_1 L_3$       & $\tilde e_R^\kappa$ & 
0.06 & 0.0055 & $R_\tau$ \\ 
\hline 
$\lambda_{23\kappa}$   & $L_2 L_3$       & $\tilde e_R^\kappa$ & 
0.06 & 0.0055 & $R_\tau$ \\ 
\hline 
$\lambda_{1\kappa 1}$  & $L_1 \lbar E_1$ & $\tilde \nu_L^\kappa$ & 
~0.37 [0.06]~ & 0.21 [0.0055] & $A_{FB}$ \\ 
\hline 
$\lambda_{2\kappa 1}$  & $L_2 \lbar E_1$ & $\tilde \nu_L^\kappa$ & 
0.25 [0.07] & 0.095 [0.0074] & $A_{FB}$ \\ 
\hline 
$\lambda_{3\kappa 1}$  & $L_3 \lbar E_1$ & $\tilde \nu_L^\kappa$ & 
0.11 [0.07] & 0.018 [0.0074] & $A_{FB}$ \\ 
\hline 
$\lambda_{1\kappa 2}$  & $L_1 \lbar E_2$ & $\tilde \nu_L^\kappa$ & 
0.25 [0.06] & 0.095 [0.0055] & $A_{FB}$ \\ 
\hline 
$\lambda_{2\kappa 2}$  & $L_2 \lbar E_2$ & $\tilde \nu_L^\kappa$ & 
0.25 [0.06] & 0.095 [0.0055] & $A_{FB}$ \\ 
\hline 
$\lambda_{3\kappa 2}$  & $L_3 \lbar E_2$ & $\tilde \nu_L^\kappa$ & 
0.25 [0.06] & 0.095 [0.0055] & $A_{FB}$ \\ 
\hline \hline
$\lambda'_{11\kappa}$  & $L_1 Q_1$       & $\tilde d_R^\kappa$ & 
0.02 & 0.0006 & $V_{ud}$ \\ 
\hline 
$\lambda'_{21\kappa}$  & $L_2 Q_1$       & $\tilde d_R^\kappa$ & 
0.06 & 0.0055 & $R_\pi$ \\ 
\hline 
$\lambda'_{31\kappa}$  & $L_3 Q_1$       & $\tilde d_R^\kappa$ & 
0.11 & 0.018  & $R_{\tau \pi}$ \\ 
\hline 
$\lambda'_{1\kappa 1}$ & $L_1 \lbar D_1$ & $\tilde q_L^\kappa$ & 
0.02 & 0.0006 & APV \\ 
\hline 
$\lambda'_{2\kappa 1}$ & $L_2 \lbar D_1$ & $\tilde q_L^\kappa$ & 
0.22 & 0.07 & $\nu_\mu$ DIS \\ 
\hline 
$\lambda'_{3\kappa 1}$ & $L_3 \lbar D_1$ & $\tilde q_L^\kappa$ & 
$\hat \lambda_{31}$ [0.52]   & 1.52\,$\hat \lambda_{31}^2$ [0.41] & 
 ``$\nu_\tau$ DIS'' \\ 
\hline 
~$\lambda'_{\kappa 11}$~ & ~$Q_1 \lbar D_1$~ & $\tilde \ell_L^\kappa$ & 
0.3 [0.11] & 0.14 [0.018] & ~$\Lambda(qqqq)$~ \\ 
\hline 
\end{tabular}
\end{center} 
\spa
\spa
\spa

In general there could also be trilinear couplings involving the
up-type quark singlet as well as couplings to scalar triplets, which
we do not discuss explicitly.  We remark that replacing the scalar
weak singlets by triplets of the same charge, while keeping the flavor
structure, only changes the sign in the doublet-doublet contraction
and yields the same effective interactions. However, a neutral triplet
may also couple to $\nu \nu$ inducing additional effective
couplings. Moreover a triplet can have flavor diagonal coupling to $L
L$, while for scalars $\lambda$ has to be antisymmetric in flavor
space. The $\Delta_L$ in left-right symmetric models is an example for
a scalar triplet with flavor diagonal couplings.  We do not consider
here the possibility of intermediate vector bosons, which will couple
to different bilinears than the scalar fields, and moreover produce a
different spin structure for the four-fermion operator.

From the definition of $\hat G_N^{A^\dagger B}$ and $\hat
G_N^{A^\dagger A}$ it follows that
\beq
\hat G_N^{A^\dagger B} = 
 \sqrt{\hat G_N^{A^\dagger A} \, \hat G_N^{B^\dagger B} }.
\eeq
In Tab.~2 we show $\hat G_N^{A^\dagger B}$ (based on this relation and
the constraints tabulated in Tab.~1) for the various $L$-violating
effective couplings that are relevant for neutrino oscillation
experiments.

\begin{center} 
Tab.~2: Experimental constraints on $L$-violating couplings \\
\spa
\spa
\begin{tabular}{|c|c|c|c|c|} 
\hline  
$A^\dagger$ & $B$ & ~$\hat G_N^{A^\dagger B}/G_F$ [\SUSYwoR]~ & 
reaction & relevant for \\
\hline \hline 
~$L_1 \lbar E_2$~ & ~$L_1 L_2$~ & 0.019 [0.0046] & 
~$\mu^+_L \to e^+_R \, \bar\nu_e \, \bar\nu_\mu$~ & LSND: DAR \\
\hline 
$L_1 \lbar E_2$ & $L_1 L_3$ & 0.023 [0.0055] &
$\mu^+_L \to e^+_R \, \bar\nu_e \, \bar\nu_\tau$ & LSND: DAR \\
\hline \hline
$L_1 \lbar D_1$ & $L_1 Q_1$ & 0.0006 &
$\nu_e \, u_L \to d_R \, e^+_R$ & LSND: ``fake'' $\bar\nu_e$ \\
\hline 
$L_2 \lbar D_1$ & $L_1 Q_1$ & 0.0067 & 
$\nu_\mu \, u_L \to d_R \, e^+_R$ &
 LSND: ``fake'' $\bar\nu_e$ \\
\hline 
$Q_1 \lbar D_1$ & $L_1 L_2$ & 0.023 [0.0083] & 
$\nu_\mu \, u_L \to d_R \, e^+_R$ &
 LSND: ``fake'' $\bar\nu_e$ \\
\hline 
$L_2 \lbar D_1$ & $L_3 Q_1$ & 0.037 &
$\nu_\mu \, u_L \to d_R \, \tau^+_R$ & 
 ~NOMAD/CHORUS: ``fake'' $\bar\nu_\tau$~ \\
\hline 
$Q_1 \lbar D_1$ & $L_2 L_3$ & 0.028 [0.010] & 
$\nu_\mu \, u_L \to d_R \, \tau^+_R$ & 
 ~NOMAD/CHORUS: ``fake'' $\bar\nu_\tau$~ \\
\hline \hline 
$L_1 \lbar E_1$ & $L_1 L_2$ & 0.028 [0.0046] &
$\nu_e \, e^-_L \to \bar\nu_\mu \, e^-_R$ & SN: $\nu_e \to \bar\nu_\mu$ \\
\hline 
$L_1 \lbar E_1$ & $L_1 L_3$ & 0.034 [0.0055] & 
$\nu_e \, e^-_L \to \bar\nu_\tau \, e^-_R$ & SN: $\nu_e \to \bar\nu_\tau$ \\
\hline 
$L_1 \lbar D_1$ & $L_1 Q_1$ & 0.0006 &  
$\nu_e \, d_L \to \bar\nu_e \, d_R$ & 
SN: $\nu_e \to \bar\nu_e$ \\
\hline 
$L_1 \lbar D_1$ & $L_2 Q_1$ & 0.0018 &
$\nu_e \, d_L \to \bar\nu_\mu \, d_R$ & 
SN: $\nu_e \to \bar\nu_\mu$ \\
\hline 
$L_1 \lbar D_1$ & $L_3 Q_1$ & 0.0033 &
$\nu_e \, d_L \to \bar\nu_\tau \, d_R$ &  
SN: $\nu_e \to \bar\nu_\tau$ \\
\hline 
$L_2 \lbar D_1$ & $L_1 Q_1$ &  0.0067 &
$\nu_e \, d_L \to \bar\nu_\mu \, d_R$ & 
SN: $\nu_e \to \bar\nu_\mu$ \\
\hline 
$L_3 \lbar D_1$ & $L_1 Q_1$ & 0.030\,$\hat \lambda_{31}$ [0.016] &
$\nu_e \, d_L \to \bar\nu_\tau \, d_R$ &  
SN: $\nu_e \to \bar\nu_\tau$ \\
\hline \hline
$L_2 \lbar E_1$ & $L_1 L_3$ & 0.023 [0.0064] &
$\nu_\mu \, e^-_L \to \bar\nu_\tau \, e^-_R$ & 
AN: $\nu_\mu \to \bar\nu_\tau$ \\
\hline
$L_3 \lbar E_1$ & $L_1 L_2$ & 0.0083 [0.0053] &
$\nu_\mu \, e^-_L \to \bar\nu_\tau \, e^-_R$ & 
AN: $\nu_\mu \to \bar\nu_\tau$ \\
\hline 
$L_2 \lbar D_1$ & $L_3 Q_1$ & 0.037 &
$\nu_\mu \, d_L \to \bar\nu_\tau \, d_R$ & 
AN: $\nu_\mu \to \bar\nu_\tau$ \\
\hline 
$L_3 \lbar D_1$ & $L_2 Q_1$ & 0.091\,$\hat \lambda_{31}$ [0.047] &
$\nu_\mu \, d_L \to \bar\nu_\tau \, d_R$ & 
AN: $\nu_\mu \to \bar\nu_\tau$ \\
\hline 
\end{tabular}
\end{center} 
\label{bilinc}
%
\spa
\spa
\spa

From Tab.~2 one can see that model-independently almost all effective
couplings for the lepton number violating operators are constrained to
be at most a few percent of $G_F$. The weakest constraints are those
involving $\lambda_{31}$, but even allowing $\lambda_{31}$ to be of
order unity implies that $G_N^{A^\dagger B} \lsim 0.1~G_F$. Imposing
SUSY we find that all of the effective couplings are constrained to be
less than one percent of $G_F$, except those involving
$\lambda'_{3\kappa1}$ which could be at most a few percent of $G_F$.

\subsection{Direct constraints}

We turn now to a discussion of additional constraints on the effective
four-fermion operators that violate total lepton number. Unlike the
bounds derived in the previous section these bounds do not depend upon
the constraints on the trilinear couplings, but apply to the
$L$-violating operator itself.

\subsubsection{Pion decays}
\label{PionDecays}

Consider the ratio between the decay rates of $\pi^+ \to e^+ \nu$
and $\pi^+ \to \mu^+ \nu$\,,
\beq \label{RpiDef}
R_\pi = {\Gamma(\pi^+ \to e^+ \nu) \over \Gamma(\pi^+ \to \mu^+ \nu)} \,.
\eeq
The measured value of this ratio~\cite{PDG},
\beq \label{RpiExpt}
R_\pi(expt) = (1.235 \pm 0.004) \times 10^{-4} \,,
\eeq
is in good agreement with the value predicted by the standard model,
including radiative corrections~\cite{DBC},
\beq \label{RpiSM}
R_\pi(SM) = (1.230 \pm 0.008) \times 10^{-4} \,.
\eeq
Consequently any non-standard contribution to either $\pi^+ \to e^+
\nu$ or $\pi^+ \to \mu^+ \nu$ is constrained to be small~\cite{KHK}.
Since the final neutrino is not detected this applies to $\pi^+$
decays with both final neutrinos and antineutrinos.  The latter case
is particularly interesting, because it allows to constrain lepton
number violating operators that induce pion decays. Note that for
these operators (unlike for lepton number conserving operators) the
decay amplitude is not suppressed by the charged lepton mass $m_\ell$
($\ell=e,\mu$), but there is an enhancement by~\cite{KHK}
\beq
f_\ell \equiv {m_\pi^2 \over m_\ell \, (m_u + m_d)} 
\eeq
with respect to the standard model currents. Therefore, when adding NP
interactions to those of the SM, to leading order in the effective
couplings $G_N^\ell \ll G_F$ of the lepton number violating operators
that induce $\pi^+ \to \ell \, \bar \nu$, the ratio in~(\ref{RpiDef})
is
\beq \label{RpiNP}
{R_\pi(SM+NP) \over R_\pi(SM)} = 
 1 + {(f_e \, G_N^e)^2 - (f_\mu \, G_N^\mu)^2 \over (V_{ud} \, G_F)^2} \,,
\eeq
where $V_{ud}$ is the CKM matrix element relevant for the SM pion
decay. Then, assuming that there are no fine-tuned cancellations
in~(\ref{RpiNP}), it follows from~(\ref{RpiExpt}) and~(\ref{RpiSM})
that
\beqs
\label{Ge}  
G_N^e   &\lsim& 3 \times 10^{-5} \, G_F \,, \\
\label{Gmu} 
G_N^\mu &\lsim& 4 \times 10^{-3} \, G_F \,.
\eeqs

We conclude that the effective couplings of all the operators in
second section of Tab.~2 that induce $\nu_\ell \, u \to d \, e^+$ must
be severely suppressed due to the bound on $G_N^e$, since they also
give rise to the lepton number violating pion decays. In particular
the model-independent bound for $(Q_1 \lbar D_1) \, (L_1 L_2)$ is
improved significantly.  Moreover, also the operator $(Q_1 \lbar D_1)
\, (L_2 L_3)$ can be constrained by the limit on $G_N^\mu$, because
the structure of the singlet bilinear
$
(L_2 L_3)_s = \nu_\mu \tau - \mu \nu_\tau \,,
$
implies that the operator obtained by exchanging the flavors of the
neutrino and the charged lepton must have the same effective coupling.
A similar argument applies to the operators $(L_\alpha \lbar D) \,
(L_\ell Q)$ ($\alpha=e, \mu, \tau$ and $\ell=e,\mu$) appearing in the
third and forth section of Tab.~2\,. Since 
$
(L_\ell Q)_s = \ell u_L - \nu_\ell d_L
$
they induce both $\nu_\ell \, d_L \to \nu_\alpha \, d_R$ and $\pi^+
\to \ell^+ \, \bar \nu_\alpha$. However the upper bounds on $G_N^\ell$
in~(\ref{Ge}) and~(\ref{Gmu}) are not useful to constrain any of the
purely leptonic operators or those involving $(Q L_3)_s$.

\subsubsection{Neutrinoless double beta decay}

The combination of the SM operator for beta decay with a new physics
operator that mediates the $L$-violating process
\beq 
\bar\nu_e \, n \to e^- \, p 
\eeq 
gives rise to neutrinoless double beta decay
($0\nu\beta\beta$)~\cite{Paes} due to the exchange of a virtual
neutrino. The crucial point is that if the leptonic current of the (if
necessary Fierz transformed) NP operator contains a right-handed
neutrino then the contribution from the neutrino propagator is
\beq
\propto P_L {q^\mu \gamma_\mu + m_\nu \over q^2 - m_\nu^2} P_R =
{q^\mu \gamma_\mu \over q^2 - m_\nu^2} \,.
\eeq
Therefore the $0\nu\beta\beta$ amplitude is proportional to $G_N \,
G_F \cdot q$, where the neutrino momentum $q$ is typically given by
the nuclear Fermi momentum $p_F \simeq 100$\,MeV~\cite{Paes}.  The
present half-life limit of the Heidelberg-Moscow experiment,
$T^{0\nu\beta\beta}_{1/2} > 1.6 \cdot 10^{25}y$~\cite{Heidelberg} then
implies severe limits on any lepton-number violating operator $(\bar u
d \bar e \nu_e^c)$. For scalar couplings one finds~\cite{Paes}
\beq
G_N[(\bar u d \bar e \nu_e^c)] \lsim 10^{-8}~G_F \,.
\eeq
For tensor couplings the constraints are even stronger.  The above
argument only applies to the operator $(L_1 \lbar D_1)\,(L_1 Q_1)$
that appears in Tab.~2, since the remaining operators also contain
leptons of the second or third generation.

Note that one cannot combine two identical NP operators that contain
one neutrino to derive a constraint on its coupling, since the
resulting $0\nu\beta\beta$ amplitude would be proportional to the
neutrino mass, which has no lower bound.

\subsubsection{Neutrino masses}

Lepton number violating operators that include two neutrinos (or two
antineutrinos) give rise to neutrino Majorana masses when closing the
external charged fermion lines by one or two loops. If the fermions in
the loop have identical flavor a contribution to the neutrino mass is
generated at one loop.

Assume that one neutrino $\nu_i$ couples to a charged fermion $f_k$
with a mass $m^f_k$ via a scalar singlet (s) or triplet (t) with
coupling $\lambda^{s,t}_{ik}$, while the second neutrino $\nu_j$
couples to another charged fermion $f_l$ via a scalar doublet (d) with
coupling $\lambda^d_{jl}$. The mixing of the equal charge components
of the two scalar fields gives rise to a $L$-violating operator as
shown in Section~\ref{formalism}. Let us consider the case where the
two charged fermions are identical, i.e. $l=k$. Then the lowest order
contribution to the neutrino mass arises at one loop (see
Fig.~\ref{one-loop-mass}). Since the charged fermion propagating in
the loop has to flip chirality a mass insertion is needed and the
neutrino mass is proportional to $m^f_k$. The (momentum dependent)
propagator for the scalar fields in the loop follows
from~(\ref{propagatorAB}) by replacing $M_{1,2}^2 \to M_{1,2}^2 -
p^2$, where $p$ is the loop momentum.  Then, the neutrino mass matrix
is given by~\cite{mass-ind}
\beqs
m^\nu_{ij} &=& i N_c \sum_k
(\lambda^{s,t}_{ik} \lambda^d_{jk} + \lambda^{s,t}_{jk} \lambda^d_{ik}) \, 
\int{d^4 p \over (2 \pi)^2} \, {m^f_k \over (m^f_k)^2 - p^2} \,
{\sin 2\phi \over 2} 
\left({1 \over M_1^2 - p^2} - {1 \over M_2^2 - p^2} \right) \\
&\simeq& N_c \sum_k \label{mass-one-loop}
{\lambda^{s,t}_{ik} \lambda^d_{jk} + \lambda^{s,t}_{jk} \lambda^d_{ik}
  \over 32 \pi^2} \,
m^f_k \, \sin 2\phi \, \ln\left({M_2^2 \over M_1^2}\right) \,,
\eeqs
where $N_c = 3~(1)$ for intermediate quarks (leptons) and the
approximation in~(\ref{mass-one-loop}) is valid for $m^f_k \ll
M_{1,2}$. 

How this result can be used to derive constraints on the effective
coupling of lepton number violating operators? Connecting two SM beta
decays with an intermediate neutrino, one induces $0\nu\beta\beta$
with an amplitude proportional to $G_F^2 \cdot m^\nu_{11}$ and the
lower bound on $T^{0\nu\beta\beta}_{1/2}$ translates into a constraint
on the $m^\nu_{11}$ entry of Majorana mass-matrix. Moreover, assuming
that two of the three mentioned neutrino problems are explained by
neutrino oscillations (implying $\Delta_{ij} \lsim 1~$eV$^2$), it
follows from unitarity that {\it all}\/ of the entries of the Majorana
mass-matrix can be at most of the order of the upper bound from the
Troitsk tritium beta-decay experiment~\cite{Troitsk} on the lightest
neutrino mass eigenstate, $m_1 < 2.5$~eV. So we have
\beq \label{mass-bound}
m^\nu_{ij} <
\left\{\matrix{0.36 \eV    & i=j=1 \cr
               \sim 3 \eV & {\rm else} }\right. \,.
\eeq
Note that if only one of the three neutrino anomalies is explained by
neutrino oscillations or if one introduces additional light (sterile)
neutrinos~\cite{sterile} the above argument for $i,j \ne 1$ does not
hold. However, since our main phenomenological motivation for
introducing $L$-violating interactions is to see whether in such a
framework all the three neutrino anomalies could be explained
simultaneously with three light neutrinos, we shall use the bounds
in~(\ref{mass-bound}) in the following.
 
Assuming that there are no significant cancellations between the
various terms that contribute to the neutrino mass
in~(\ref{mass-one-loop}) it follows from~(\ref{mass-bound}) that
\beq 
\lambda^{s,t}_{ik} \lambda^d_{jk}
\, \sin 2\phi \, \ln\left({M_2^2 \over M_1^2}\right) \lsim
\left({{\rm MeV} \over N_c m^f_k}\right) \cdot
\left\{\matrix{6.3 \times 10^{-5} & i=j=1 \cr
               9.5 \times 10^{-4} & {\rm else} }\right. \,.  
\eeq
This implies that the effective coupling of the $L$-violating operator
satisfies:
\beqs
G_N[(\lbar{f_{kL}} \, \nu_i) \, (f_{kR} \, \nu_j^c)] &=&
{\lambda^{s,t}_{ik} \lambda^d_{jk} \over 8 \sqrt2} \, \sin 2\phi
\left( {1 \over M_1^2} - {1 \over M_2^2} \right) \\ 
&\lsim& f(M_2/M_1) \, \left({{\rm MeV} \over N_c m^f_k}\right) \cdot
\left\{\matrix{1.9 \times 10^{-4}~G_F & i=j=1 \cr
               2.9 \times 10^{-3}~G_F & {\rm else} }\right. \,,  
\label{nu-bound}  
\eeqs
where
\beq
f(x) \equiv {1-x^{-2} \over \ln x^2} < 1 ~~~{\rm for}~x > 1 \,,
\eeq 
and we have set the lower mass $M_1 = 50$~GeV to its minimal value.
We learn that the above constraints on $G_N$ in many cases are
stronger than those listed in the third and fourth section of
Tab.~2\,. In particular, all the effective couplings for lepton number
violating scattering off electrons and quarks are at most of the order
$6 \times 10^{-3}~G_F$ and a few$\,\times 10^{-4}~G_F$, respectively.

\section{Lepton number violating interactions and neutrino oscillation 
  experiments}
\label{exps}

Having introduced and motivated the $L$-violating interactions induced
by scalar mixing, we turn now to a systematic survey of those
interactions that are relevant to the terrestrial, solar and
atmospheric neutrino experiments.

\subsection{Terrestrial neutrino experiments}

The LSND collaboration has reported a positive signal in two different
appearance channels. The first analysis~\cite{LSND-DAR} uses $\bar
\nu_\mu$'s from muon decay at rest (DAR) and searches for $\bar
\nu_e$'s via inverse beta decay. The observed excess of $\bar \nu_e$
events corresponds to an average transition probability
of~\cite{LSND-DAR}
\beq \label{LSNDdata}
P(\bar \nu_\mu \to \bar \nu_e)=(3.1^{+1.1}_{-1.0} \pm 0.5) \times 10^{-3}.
\eeq
Explaining this result in terms neutrino oscillations, requires
$\Delta m^2$ and $\sin^2 2\theta$ in the range indicated in Fig.~3 of
Ref.~\cite{LSND-DAR}. Taking into account the restrictions from the
null results of other experiments, the preferred values of the
neutrino parameters are $\Delta m^2 \approx 2~\eV^2$ and $\sin^2
2\theta \approx 2 \times 10^{-3}$ and the {\it lower} limit on $\Delta
m^2$ for the neutrino oscillation solution is given by
\beq \label{LSNDrange} 
\Delta m^2 > 0.3~\eV^2 \,.
\eeq 
The second analysis~\cite{LSND-DIF} uses $\nu_\mu$'s from pion decay
in flight (DIF) and searches for $\nu_e$'s via the $\nu_e \, C \to e^-
\, X$ inclusive reaction. Again a positive signal with a transition
probability $P(\nu_\mu \to \nu_e)$ similar to the one
in~(\ref{LSNDdata}), but with less statistical significance, has been
reported.

Besides the orthodox neutrino oscillation hypothesis it has been
proposed that the LSND signals could be due to non-standard neutrino
interactions~\cite{Grossman,Herczeg,Johnson}. Assuming that the result
in~(\ref{LSNDdata}) is due to New Physics interactions of strength
$G_N^\nu$ (while there is no significant contribution from neutrino
oscillations) the appearance probability is given by~\cite{Grossman}
\beq \label{xzero}
P(e^+) = \left|{G_N^\nu \over G_F}\right|^2 \,.
\eeq
From eqs.~(\ref{LSNDdata}) and~(\ref{xzero}) we learn that, in order
to explain the LSND result, the effective NP coupling should satisfy
\beq \label{Grequired}
G_N^\nu > 4.0 \times 10^{-2}~G_F 
\eeq
at the 90\% confidence level (CL).  In Ref.~\cite{BG} it has been
shown that lepton flavor violating neutrino interactions cannot
satisfy the condition~(\ref{Grequired}) even if one allows for maximal
$SU(2)_L$ breaking effects.

Here we investigate whether lepton number violating interactions could
be large enough to be relevant for the LSND results.  As we already
mentioned in the Introduction the anomalous $L$-violating
decays~(\ref{amdlv}) could in principle be a possible source for the
$\bar \nu_e$'s in the DAR channel of LSND.  From the first section of
Tab.~2 it follows that these decays can be mediated by a scalar
doublet that couples to $A^\dagger=L_1 \lbar E_2$ whose charged
component mixes with a scalar singlet that couples to $B=L_1 L_\ell$.
(Here $\ell=\mu, \tau$, but for a scalar triplet also $\ell=e$ is
possible.) Comparing the model-independent bounds for the effective
couplings $G_N^{A^\dagger B} \lsim 0.02\,G_F$ with the required
effective coupling strength in~(\ref{Grequired}) we conclude that in
the $SU(2)_L$ symmetric case $G_N^{A^\dagger B}$ is too small to
explain the LSND DAR result.  However, while $SU(2)_L$ breaking
effects cannot be large, an enhancement by a factor of two, which is
required to satisfy~(\ref{Grequired}), is indeed conceivable.  Thus we
cannot rule out in a model independent way that the lepton number
violating decays in~(\ref{amdlv}) are the source of the LSND anomaly.
However, moving to the explicit framework of \SUSYwoR\ the constraints
on $G_N^{A^\dagger B}$ are stronger by a factor of four implying that
even with maximal $SU(2)_L$ breaking one cannot
fulfill~(\ref{Grequired}), unless one allows for some fine-tuned
cancellations.

It is interesting to ask whether the $L$-violating interactions
\beq \label{nuplv}
\nu_\ell \, p \to e^+ \, n \,.
\eeq
could provide an alternative explanation for the DAR signal. As one
can see from the second section of Tab.~2 the processes
in~(\ref{nuplv}) can be induced by scalar mixing if either
$A^\dagger=L_\ell \lbar D_1$ to $B=Q_1 L_1$ or ${A'}^\dagger=Q_1 \lbar
D_1$ to $B'=L_1 L_\ell$.  While the scalar fields coupling to $A$ and
$A'$ have to be doublets, those coupling to $B$ and $B'$ could be
either singlets or triplets of $SU(2)_L$. (Note that if $L_1 L_\ell$
couples to a singlet then this excludes $\ell=e$ due to the
antisymmetry of the singlet contraction.) However, as we have noted in
Section~\ref{PionDecays}, any operator that induces the reaction
in~(\ref{nuplv}) also necessarily gives rise to lepton number
violating pion decays. Thus, the stringent constraints in~(\ref{Ge})
and~(\ref{Gmu}) apply, unless one is willing to allow for a fine
tuned cancellation in~(\ref{RpiNP}) by setting $G_N^e/m_e =
G_N^\mu/m_\mu$. But even in this case according to the bound in Tab.~2
the effective coupling $G_N^{A^\dagger B}$ of the operator $A^\dagger
B$ is much too small to satisfy~(\ref{Grequired}).  Using only the
bounds from the trilinear couplings in Tab.~2 $G_N^{{A'}^\dagger B'}$
could be consistent with~(\ref{Grequired}) provided that there is an
enhancement from $SU(2)_L$ breaking effects by a factor of two. The
corresponding bound within \SUSYwoR\ is only stronger by a factor of
three. Although it is rather unlikely, we cannot rule out completely
that the lepton number violating reactions in~(\ref{nuplv}) play a
role for the LSND DAR result.

We note that lepton number violating pion decays $\pi^+ \to \ell^+ \,
\bar\nu_e$ cannot be responsible for the $\bar\nu_e$'s observed by
LSND, even though they are not helicity suppressed. First, according
to~(\ref{RpiExpt}) the BR for $\ell=e$ is {\it measured} to be too
small.  Second, for $\ell=\mu$ the kinetic energy of the final
$\bar\nu_e$ is at most 34~MeV, which is below the threshold energy of
the LSND DAR analysis.

As concerns the LSND DIF channel an interpretation of this anomaly in
terms of $L$-violating interaction is less attractive for the
following reason: The presence of additional $L$-violating pion decays
of the form $\pi^+ \to \mu^+ \, \bar\nu_\ell$ cannot produce the
observed $\nu_e$'s. Likewise $L$-violating interactions in the
detection process could only imply that neutrons capture antineutrinos
which are absent in the SM pion DIF.  Hence the (generically
suppressed) $L$-violation processes would be required for both the
neutrino production and detection, ruling out this scenario as an
explanation for the LSND DIF signal.

We note that the KARMEN experiment~\cite{KARMEN}, which uses the same
detection processes as LSND has found no evidence for neutrino flavor
transitions ruling out a transition probability as in~(\ref{LSNDdata})
at 90\%~CL. In general this situation somewhat favors an explanation
of the LSND anomaly in terms of ``standard'' neutrino oscillations,
since due to the different baselines there is still as small region in
the $\Delta m^2 -\sin^2 2\theta$ plane consistent with both
experiments, while for new physics reactions as the source of the LSND
anomalies KARMEN should observe the same transition probabilities.
Still, the bound from KARMEN gains from the fact that less events were
observed than expected from the background, so for conclusive evidence
we will have to wait for the upcoming MiniBooNE
experiment~\cite{MiniBooNE} (see also Section~\ref{terfuture}).
 
A different search for neutrino oscillations has been performed by the
CHORUS~\cite{CHORUS} and NOMAD~\cite{NOMAD} experiments at CERN
looking for $\nu_\mu \to \nu_\tau$ oscillations transitions. In the
absence of neutrino flavor transitions the relative flavor composition
of the neutrino beam is predicted to be $\nu_\mu : \bar\nu_\mu :
\nu_e : \bar\nu_e = 1.00 : 0.061 : 0.0094 : 0.0024$ with a negligible
($\simeq 10^{-7}$) contamination of tau neutrinos. The search for
$\nu_\tau$ is based on charged current tau production with subsequent
detection of the various tau decay modes. Both experiments have found
no indication for $\nu_\mu \to \nu_\tau$ oscillation. The upper bound
from NOMAD~\cite{NOMAD} on the transition probability is
\beq \label{PNomad}
P(\nu_\mu \to \nu_\tau) < 0.6 \times 10^{-3} ~~(90\%~CL) \,.
\eeq
Since also the observed $\tau^+$ events are in agreement with the
estimated background a similar bound as in~(\ref{PNomad}) applies to
$P(\nu_\mu \to \bar\nu_\tau)$.  The production of $\tau^+$'s could
also be induced by the lepton number violating reaction
\beq
\nu_\mu \, p \to \tau^+ \, n \,,
\eeq
which would result from the operators $(L_2 \lbar D_1) \, (L_3 Q_1)$
or $(Q_1 \lbar D_1) \, (L_2 L_3)$ that appear in Tab.~2\,. It is
interesting to note that the upper bounds we obtained for the
effective coupling of these operators in Section~\ref{constraints}
(see Tab.~2) are of the same order [$(\hat G_N/G_F)^2 \sim 10^{-3}$]
as the experimental constraint from NOMAD (and CHORUS). Unfortunately
the proposed TOSCA experiment~\cite{TOSCA} that would have been
sensitive to a transition probability as small as $\sim 10^{-5}$ has
been rejected.

\subsection{Solar neutrino experiments}
\label{SN}

The long standing solar neutrino puzzle~\cite{SNP} is now confirmed by
five experiments using three different experimental techniques and
thus probing different neutrino energy ranges. All these experiments
observe a solar neutrino flux that is smaller than expected. The most
plausible solution is that the neutrinos are massive and there is
mixing in the lepton sector. Then neutrino oscillations can explain
the deficit of observed neutrinos with respect to the Standard Solar
Model. In the case of matter-enhanced neutrino oscillations, the
famous MSW effect provides an elegant solution~\cite{SNP} to the solar
neutrino problem with $\Delta_{SN}$ as given in~(\ref{DeltaSN}).

Several authors have studied alternative solutions to the solar
neutrino problem with and without neutrino
masses~\cite{SN-NP,Bergmann}. In the scenario with massive neutrinos
$\Delta_{SN}$ is still required to be of the same order as
in~(\ref{DeltaSN}). However, the vacuum mixing can be vanishingly
small~\cite{Bergmann}, when the effective mixing is dominantly induced
by the flavor changing neutrino scattering
\beq \label{FC}
\nu_e \, f \to \nu_\ell \, f,
\eeq
where $\ell=\mu, \tau$ and $f=e, u, d$.

For the scenario without neutrino masses additional non-universal
flavor diagonal interactions
\beq \label{FD}
\nu_\ell \, f \to \nu_\ell \, f,
\eeq
are required. For both scenarios the effective couplings in~(\ref{FC})
$G_{e \ell}^f$ have to be of the order of a few percent, while for the
second scenario the difference between the effective couplings
in~(\ref{FD}) $G_{\ell \ell}^f-G_{ee}^f$ has to be in the narrow
interval $[0.50\,G_F, 0.77\,G_F]$ ($[0.40\,G_F, 0.46\,G_F]$) for
$f=d~(u)$ to allow for a resonant neutrino conversion. This requires
rather large non-universal flavor diagonal couplings, which is ruled
out for $\ell=\mu$~\cite{Bea}.

It is interesting to ask whether also the $L$-violating neutrino
scattering
\beq \label{FClnv}
\nu_e \, f \to \bar\nu_\ell \, f,
\eeq
where $\ell=\mu, \tau$ and $f=e, u, d$ in combination with either
massive neutrinos (but negligible mixing) or additional non-universal
flavor diagonal interactions of the type
\beq \label{FDlnv}
\bar\nu_\ell \, f \to \bar\nu_\ell \, f,
\eeq
could give rise to matter-induced $\nu_e-\bar\nu_\ell$ neutrino
oscillation that provide an alternative solution to the solar neutrino
problem. Note that the $L_\ell$ and $L$ conserving interactions
in~(\ref{FD}) and~(\ref{FDlnv}) are related by crossing symmetry. So
their effective couplings are subject to the same bound.

As we have seen in Section~\ref{formalism} lepton number violating
reactions as in~(\ref{FClnv}) require mixing between intermediate
bosons with different $SU(2)_L$ transformations.  The third section of
Tab.~2 contains various combinations of bilinears that when coupled to
each other by scalars that mix can induce the $L$-violating neutrino
scattering off a fermion as in~(\ref{FClnv}).  The total lepton number
will only be violated if the two bilinears $A_{q_f}=\bar\nu f_A$ and
$B_{q_f}=\nu f_B$ contain charged fermions $f_A$ and $f_B$ that belong
to different presentations of $SU(2)_L$.  Consequently, independent of
the details of the model, the reordered four-fermion operator that
induces the effective neutrino potential
\beq \label{Hint}
{\cal H}_{\rm int} = 
{G_F \over\sqrt{2}} 
\sum_{a=S,P,T} (\lbar{\nu^c}\,\Gamma^a\,\nu)\, 
\left[\lbar \psi_f\,\Gamma_a\, (g_a + g'_a
\gamma^5)\,\psi_f\right]\, + {\rm h.c.}\,, 
\eeq
can only contain scalar ($\Gamma^S=I$), pseudo-scalar
($\Gamma^S=\gamma_5$) or tensor ($\Gamma^T=\sigma^{\mu\nu}$)
couplings. (Axial)vector couplings are not possible, since they couple
between fermions of the opposite chirality.

To be explicit consider the example within \SUSYwoR\ where the $q=1/3$
component of $A^\dagger = L_3 \lbar D_1$ may couple to the weak
singlet $B = L_1 Q_1$ if there is $\tilde b_L - \tilde b_R$ mixing.
The resulting four-fermion operator is:
\beq \label{fourfermi}
{\cal H}_{\rm int} = {\lambda'_{331} \lambda'_{313} \over M_{AB}^2} 
(\lbar{d_R} \, \nu_\tau) \, (\lbar{\nu_e^c} \, d_L) = 
- {\lambda'_{331} \lambda'_{313} \over M_{AB}^2} 
\left[
\half(\lbar{\nu_e^c} \, \nu_\tau) \, (\lbar{d_R} \, d_L) + 
\eighth
(\lbar{\nu_e^c} \, \sigma_{\mu \nu} \, \nu_\tau) \, 
(\lbar{d_R} \, \sigma^{\mu \nu} \, d_L)
\right] \, .
\eeq 
The question is then whether scalar and tensor couplings can affect
the neutrino propagation in dense matter significantly. The bounds
from the neutrino masses in~(\ref{nu-bound}) indicate that the
relevant effective couplings $G_N$ are less than $10^{-2(-3)}~G_F$ for
neutrino scattering of electrons (quarks). This constraint could be
evaded if we only accept one measurement for mass squared difference
$\Delta m^2$ and allow for one mass eigenstate much heavier than
$2.5$~eV. However, we still have the bounds from the trilinear
couplings (see Tab.~2), which imply that $G_N$ is at most at the few
percent level. Then the lepton number violating interactions could
only affect the standard MSW oscillations if the averaged matrix
element of the background fermion current in~(\ref{Hint}) is of
similar order as the one from the SM weak current~\cite{Bergmann}.  In
Ref.~\cite{BGN} it has been shown that for (pseudo)scalar interactions
the effective neutrino potential that is induced by~(\ref{Hint}) is
proportional to the ratio of the neutrino mass and the characteristic
fermion energy. Thus (pseudo)scalar couplings in~(\ref{Hint}) are not
relevant for matter induced neutrino oscillations in the
Sun. Moreover, it has been pointed out that transverse tensor
couplings are not suppressed by the neutrino mass~\cite{BGN}. However,
in this case the effective neutrino potential is proportional to the
average (transverse) polarization of the background matter. Since the
polarization in the solar interior due to the magnetic field is
expected to be tiny~\cite{BGN}, we conclude that the $L$ violating
neutrino scattering in~(\ref{FClnv}) is not relevant for the solar
neutrino problem.

\subsection{Atmospheric neutrino experiments}

Several experiments have observed an anomalous ratio between the
atmospheric muon neutrino and electron neutrino fluxes~\cite{ANP}.
This atmospheric neutrino problem has recently been confirmed by the
Super-Kamiokande high statistics data~\cite{SKAN}. Explaining this
result in terms of ``standard'' neutrino oscillations~\cite{ANP}
requires a mass squared difference $\Delta_{AN}$ as shown
in~(\ref{DeltaAN}).

Recently an alternative solution to the atmospheric neutrino anomaly
based on new neutrino interactions was proposed~\cite{AN-NP}.  The
suggested scenario is similar to the one we discussed previously for
the solar neutrino, but for $\nu_\mu-\nu_\tau$ oscillations.  Even if
neutrino masses are negligible the effective mixing between the flavor
eigenstates could in principle be induced by the flavor changing
neutrino scattering
\beq \label{FCAN}
n_\mu \, f \to n_\tau \, f,
\eeq
where $n=\nu, \bar\nu$ and $f=e, u, d$, in combination with
non-universal flavor diagonal interactions
\beq \label{FDAN}
n_\ell \, f \to n_\ell \, f,
\eeq
with $\ell=\mu, \tau$. According to~\cite{AN-NP} the effective
couplings $G_{\mu \tau}^f$ for~(\ref{FCAN}) and the difference between
the effective couplings for~(\ref{FDAN}), $G_{\tau \tau}^f-G_{\mu
  \mu}^f$, have to be both of order $0.1~G_F$. As has been shown in
Ref.~\cite{BGP} $G_{\mu \tau}^f$ is constrained by electroweak
precision data to be at most at the few percent level ruling out such
an explanation, unless one allows for some fine-tuned cancellations.

However, in view of the bounds in the fourth section of Tab.~2 on the
effective couplings for the lepton number violating neutrino
scattering
\beq \label{FClnvAN}
\nu_\mu \, f \to \bar\nu_\tau \, f \mbox{~~~~and~~~~}
\bar\nu_\mu \, f \to \nu_\tau \, f, 
\eeq
one might wonder whether such interactions could offer an alternative
mechanism to solve the atmospheric neutrino anomaly. Although the
effective coupling for $(L_3 \lbar D_1) \, (L_2 Q_1)$ could be of
order $0.1 \, G_F$, such an explanation faces the same problem that we
encountered in the discussion of solar neutrinos.  Namely, the
inherent change of the chirality of the background fermions restricts
the couplings to be of scalar or tensor type.  Consequently the
effective neutrino potential is suppressed by either the neutrino mass
or the average polarization. Thus we conclude that lepton number
violating interactions do not effect atmospheric neutrinos that
propagate through earth matter.

\subsection{Future terrestrial neutrino oscillation experiments}
\label{terfuture}

The fundamental difference between neutrino transitions induced by new
interactions and ``standard'' vacuum neutrino oscillations due to
non-vanishing neutrino masses and mixing is that only the latter have
a non-trivial $L/E$ (distance over energy) dependence if the neutrinos
propagate in vacuum. Flavor changing neutrino interactions that
conserve total lepton number in principle can induce matter-induced
neutrino oscillations that are distance dependent. Among laboratory
neutrinos matter effects are only relevant for long baseline
experiments, where the neutrinos propagate through the earth
mantle~\cite{LBmatter}. However, for $\nu_\mu \to \nu_\tau$
transitions the flavor changing parameter $\epsilon$ has be of order
unity~\cite{Gago} to be relevant for the K2K~\cite{K2K} and
MINOS~\cite{MINOS} long baseline neutrino experiments, which is
inconsistent with the model-independent bounds presented in
Refs.~\cite{BGP}. Also for $\nu_e \to \nu_\mu (\nu_\tau)$ transitions
$\epsilon$ is at most of order $10^{-5} (10^{-2})$~\cite{Bea} implying
that earth effects will not probe the flavor changing interactions in
the upcoming long baseline detectors.

As we pointed out in our discussion of solar neutrinos in
Section~\ref{SN} matter effects on the neutrino propagation due to
lepton number violating interactions are even less significant due to
the suppression from the neutrino mass or the background polarization.
Therefore the dominant impact on terrestrial neutrino experiments from
new interactions comes from the modification of the relevant detection
and/or production processes, like the reaction in~(\ref{amdlv}) that
we discussed for LSND. This fact allows us to distinguish the proposed
solution of LSND in terms of lepton number violating interactions
without any theoretical assumptions, just on the basis of the
experimental observations. In the future a number of terrestrial
neutrino experiments with different baselines will try to clarify the
nature of neutrino oscillations. Should a certain neutrino transition
channel maintain a distance independent contribution (beyond the
trivial decrease of the flux inverse to the distance squared) this
would signal non-standard neutrino interaction. In the following we
discuss briefly some of the upcoming experiments and their potential
to observe lepton number violating interactions.

The MiniBooNE experiment~\cite{MiniBooNE} at Fermilab is designed to
confirm (or refute) the $\nu_e-\nu_\mu$ oscillations signals observed
at LSND by searching for $\nu_\mu \to \nu_e$ transitions with an
expected sensitivity to $P(\nu_\mu \to \nu_e) > 2 \cdot 10^{-4}$ at
90\%~CL. Also a search for $\bar\nu_\mu \to \bar\nu_e$ seems
feasible even though $\bar\nu_\mu$'s are produced less copiously (by
a factor $\sim 0.2$). The important feature of MiniBooNE is its small
background of $\nu_e$ and $\bar\nu_e$, which allows to search also
for the lepton number violating transitions $\nu_\mu \to \bar\nu_e$
and $\bar\nu_\mu \to \nu_e$ with a sensitivity of a few times
$10^{-4}$. Therefore it might be possible to probe the operator $(Q_1
\lbar D_1) \, (L_1 L_2)$ at the level of its model-independent bound
obtained in Section~\ref{constraints} (c.f. Tab.~2).

Several long baseline experiments~\cite{Zuber} will search for
neutrino transitions in particular aiming at the atmospheric region of
evidence for $\nu_\mu \to \mu_\tau$ oscillations. While the long
baseline allows these experiments to explore small mass-squared
differences $\Delta m^2 \gsim 10^{-3} \eV^2$, the neutrino flux at the
detector of the upcoming experiments is rather small yielding at most
a few hundred events per year. Therefore we do not expect that these
experiments could probe lepton number violating interactions anywhere
close to the bounds obtained in Section~\ref{constraints} (c.f.
Tab.~2).  However it might be possible to obtain interesting
information from a second detector very close to the neutrino source,
which is supposed to study the initial neutrino beam.

\subsection{Supernova neutrinos}

While the effects from lepton number violating neutrino scattering as
in~(\ref{FClnv}) and~(\ref{FClnvAN}) are negligible for solar and
atmospheric neutrinos, these reactions could be relevant for the
neutrinos emerging from a supernova explosion. Here tensor
interactions could affect the neutrino propagation provided that there
is a very large magnetic field $B \sim 10^{16}\,$G which induces a
large polarization $\vev{\lambda_f} \simeq
10^{-2}-10^{-1}$~\cite{BGN}.

\section{Conclusions}
\label{conclusions}

We have presented a comprehensive analysis of lepton number violating
neutrino interactions. $L$-violating four-fermion operators involving
one or two neutrinos are induced by heavy boson exchange, if there is
mixing between the equal charge components of a doublet and a singlet
(or a triplet) of $SU(2)_L$.

As an example we have discussed \SUSYwoR, where such operators are
induced by the mixing of ``right-handed'' sfermions that are $SU(2)_L$
singlets with the ``left-handed'' sfermions that are $SU(2)_L$
doublets.

We have studied four approaches to constrain the $L$-violating
operators in a model-independent framework:
\begin{enumerate} 
\item Constraints from the trilinear couplings,
\item Universality in pion decays,
\item Neutrinoless double beta decay,
\item Loop-induced neutrino masses.
\end{enumerate}
Any non-vanishing coupling between any fermionic bilinear and a scalar
field induces an effective four-fermion operator containing the
bilinear and its hermitian conjugate with an effective coupling
proportional to the square of the trilinear coupling over the scalar
mass. Combining the upper bounds on such effective operators one can
derive constraints on the lepton number violating operators that
consist of two different bilinears.

The measured ratio between the BR for pion decays with final electrons
and those with final muons is in good agreement with the value
predicted by the SM. This implies severe constraints on NP contribution
to these pion decays. Since the final neutrinos are not observed this
includes lepton number violating pion decays. Moreover, since these
decays are not helicity suppressed, there is a significant enhancement
from the hadronic matrix element, which gives rise to stringent limits
on the relevant effective couplings.

Any $L$-violating four-fermion operator that can be combined with the
SM operator responsible for beta decay in order to induce neutrinoless
double beta decay is severely constrained by the experimental limit on
this process. Unlike for the ``standard'' neutrinoless double beta
decay, where the lepton number violation is induced by the neutrino
Majorana mass, in the New Physics case, $L$ is broken by the operator
and the intermediate neutrino can contribute by its momentum rather
than its mass.

\begin{figure}[htb]
\LARGE
\begin{center}
\mbox{\epsfig{figure=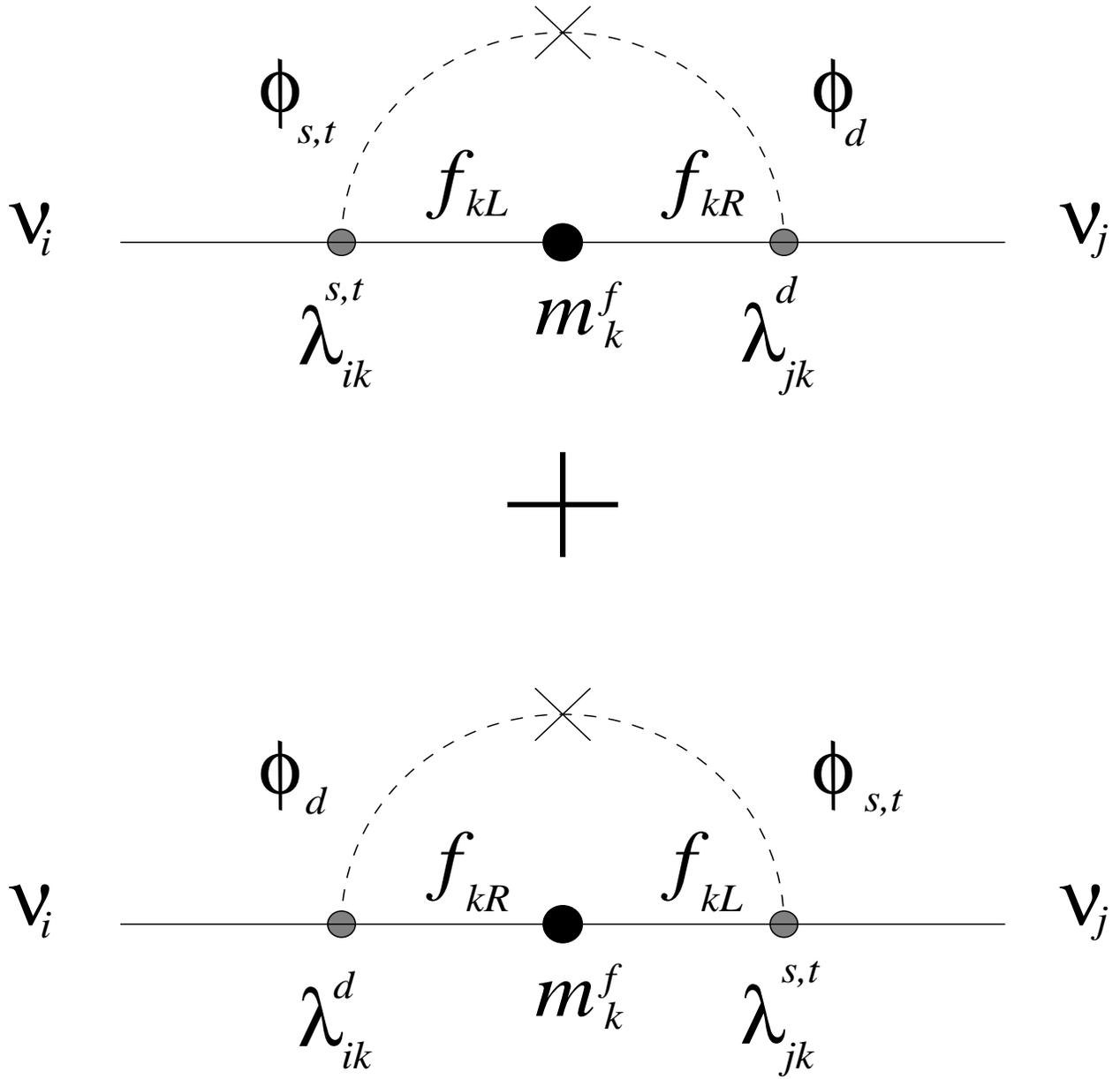,angle=0,width=16.6cm,height=16.2cm}}
\newline
\end{center}
\caption{Feynman diagram for neutrino Majorana mass term $m^\nu_{ij}$
         induced at one loop.}
\label{one-loop-mass}
\end{figure}

Lepton number violating operators containing two neutrinos and two
charged fermions of identical flavor (but opposite chirality) induce
neutrino Majorana masses at one loop when the external fermions lines
are connected by a fermion propagator (see Fig.~\ref{one-loop-mass}).
Using the upper bound on the lightest neutrino mass-eigenstate from
the Troitsk tritium experiment and assuming a three neutrino framework
with mass-splittings not larger than $\sim 1~\eV$, we derive stringent
constraints on the relevant lepton number violating operators.

Our constraints lead to the following conclusions:
\begin{itemize} 
\item Lepton number violating neutrino scattering $\nu_\alpha \, f \to
  \bar \nu_\beta \, f$ off matter fermions $f=e,u,d$ are severely
  suppressed by the bounds from neutrino masses and do not play a role
  for the present solar and atmospheric neutrino anomalies. Since any
  effect due to such interactions would require also a polarized
  background, it could -- at best -- be relevant for supernova
  neutrino oscillations.
\item Model-independent considerations show that only the operators
  $(L_1 \lbar E_2) \, (L_1 L_{2,3})$ (inducing the anomalous muon
  decay $\mu^+_L \to e^+_R \, \bar\nu_e \, \bar\nu_{\mu,\tau}$) could
  have an effective coupling at the few percent level (of $G_F$) and
  thus might be significant for the LSND anomaly.  Lepton number
  violating neutrino capture by protons is severely constrained by the
  data on pion decays, and not relevant for LSND, unless one is
  willing to accept some fine-tuned cancellations. Within \SUSYwoR\ 
  the relevant constraints are stronger by a factor of four, and an
  explanation of the LSND DAR data via lepton number violating
  interactions is inconsistent with the upper bound on the maximal
  $SU(2)_L$ breaking.
\end{itemize}

A solution of LSND in terms of New Physics is attractive since this
way the solar and the atmospheric neutrino anomalies could be
explained via standard neutrino oscillations avoiding the introduction
of a sterile neutrino. However, this interpretation seems to be
somewhat disfavored by the confirmation of the LSND anomaly in the DIF
data and by the null signal of KARMEN.

Future terrestrial neutrino oscillation experiments that are sensitive
to the ``neutrino transition'' probability at the level of $10^{-4}$
could observe lepton number violating neutrino interactions. Any
signal that does not depend on the baseline is a potential candidate
for new neutrino interactions. However, the present solar and
atmospheric neutrino anomalies are not (significantly)
``contaminated'' by such interactions.

\acknowledgements

We thank Y. Grossman, St. Kolb, W.C. Louis and Y. Nir for useful
discussions and comments. One of us (SB) would like to thank the
Max--Planck--Institut f\"ur Kernphysik in Heidelberg, where part of
this work was done, for the hospitality.



\end{document}